# Nature of Cr Segregation in FCC to BCC quenched Fe-12Cr Alloy, with Post-Quenching Heat Treatment: A Positron Annihilation Study

S. Hari Babu[1, 2, 3 #], V. R. Reddy[1]

[1] *UGC-DAE CSR,University Campus, Khandwa Road, Indore-452017, India.*
[2] *Materials Science Group, Indira Gandhi Center for Atomic Research, Kalpakkam-603102, India.*
[3] *Department of Physics, Vishnu Women's Univeristy, Bhimavaram – 534202, India*

[#]Email: haribabusata@gmail.com

**Abstract:** The Fe-Cr binary alloy is a model system for ferrtitic/martensitic steels. Cr clustering instigated loss of ductility, is one of the influential factors to deteriorate the alloys, and its severity depends on Cr concentration. This study aims at understanding (i) the nature of Cr precipitation in Fe-12Cr alloy using positron lifetime spectroscopy, transmission electron microscopy and Mössbauer spectroscopy studies and (ii) comparing the precipitation concentration and size with respect to Fe-9Cr alloy. The quenched alloys (from the high temperature FCC phase) are found to contain a high density of dislocations along with the sub-grain structure which lead to Cr rich σ-phase precipitation during post-quenching heat treatment at 748 K, while the alloy prepared with no dislocations and the sub-grain structure shows no precipitation, except at grain boundaries, consistent with observations in Fe-9Cr alloys. Selected area diffraction confirmed that the precipitates are tetragonal intermetallic σ-phase, contradicting the majority of the literature on Fe-Cr ferritic alloys which reported it as Cr rich BCC α'-phase. Mossbauer spectroscopy studies showed signatures of Cr depletion from the rest of the matrix, indicating that the precipitates are Cr rich compared to the alloy composition. The kinetics of Cr precipitation is found to be higher in the Fe-12Cr alloy compared to the Fe-9Cr alloy. The average precipitate size is observed to be higher and the number density is lower in the Fe-12Cr alloy compared to the Fe-9Cr alloy. In addition, the experimental positron lifetime in defect-free lattice of Fe-(9, 12)Cr alloys is reported for the first time with substantial evidence.

## Introduction

Ferritic/martensitic alloys are considered to be candidate structural materials for the first wall and breeding blanket in nuclear fusion reactors [1] and in the core of future fast reactors [2]. They are also considered to be a potential structural material for spallation targets of Accelerator Driven Systems (ADS), which envisage intense neutron fluxes produced by proton beams operating at GeV energy with MW range power. They are potential candidates due to their (i) swelling resistance under intense neutron irradiation [3] (ii) impressive thermal conductivity [4, 5] and (iii) flexibility to attain reduced activation by appropriate choice of alloying elements and their compositions [6].

In ferritic/martensitic alloys, embrittlement is one of the main issues affecting the properties of ferritic/martensitic alloys containing Cr, and the degree of embrittlement depends non-linearly on Cr concentration. Cr segregation is found to be one of the influential factors that affect embrittlement. Thermal aging around 748 K or/and irradiation is known to cause embrittlement (known as 475 °C embrittlement) in ferritic steels with Cr concentration greater than 12% [7]. So far, it is qualitatively understood that the hardening is due to Cr segregation and it is claimed to be in the form of coherent, Cr rich α' phase, as described in review of Malerba et. al. [8]. However, Fe and Cr lattice parameter values are very close (i.e., 2.87 Å and 2.91 Å, respectively) and the lattice parameter of BCC Fe-Cr alloy as well. Hence, theory does not support that the magnitude of increase in hardness is due to co-existence of coherent BCC(α') phase in BCC(α) [8]. This necessitated study of Fe-Cr binary model alloys to identify the precise nature of segregated phase. Also, Fe-Cr alloys are academically interesting due to the ambiguity pertaining to equilibrium phase boundaries in Fe rich side of phase diagram [9].

In Fe-Cr binary alloys, it is observed that the formation of vacancy clusters, upon irradiation, is hindered as Cr concentration is increased from 2.5 to 12 % [10]. This indicates that embrittlement due to irradiation, which is governed by the formation of voids, decreases with increase in Cr concentration, leaving Cr segregation as major factor for embrittlement in alloys with Cr concentration above 9 %. For alloys around 9-12 % Cr, there are various reports on neutron irradiation induced short range ordering [11], Cr segregation [12, 13]. There are both experimental and theoretical studies evidencing Cr segregation upon electron irradiation [14, 15, 16], showing a tendency to resist the formation of a Cr-vacancy complex [16]. Simulation work showed a tendency for clustering even for pure thermal aging in alloys with Cr concentration above 12% [17, 18, 19], a resultant of motivation to explain the emrittlement in high chromium steels. Contrary, in Fe-Cr binary alloys with Cr concentration around 10%, it was not experimentally proven for very long time. However, positron lifetime studies showed Cr rich intermetallic precipitation in pure Fe-9Cr binary alloy with an initial non-equilibrium microstructure similar to martensite, a resultant of quenching across FCC-BCC phase transition, was reported previously [20]. It is understood that the quenched-in defects, mostly dislocations and sub-grain boundaries, have a major role in Cr precipitation. Alloy free from defects shows no evidence of precipitation in positron annihilation studies. However, electron microscopy studies were not carried out for defect-free alloy [20].

The Fe-Cr phase diagram contains a high temperature FCC ($\gamma$) loop between 1119 K and 1667 K for Cr concentration below approximately 11.2 wt% in, otherwise, BCC ($\alpha$) over the whole temperature range below melting temperature. A narrow ($\alpha + \gamma$) range that exists between 11.2 and 13.4 wt% is worth noticing and is of interest to this study, as the concentration of the samples studied in this work position in this range. Hence, Fe-12Cr alloy contains a fraction of $\alpha$-phase at 1423 K which will not undergo phase transformation during quenching, while Fe-9Cr alloy, reported earlier, exists in single phase at 1423 K.

Positron annihilation spectroscopy (PAS), having sensitivity to the variation in defects that are associated with changes in the microstructure, can serve as an effective tool in finding out microstructural changes [21, 22]. PAS is extensively used as a tool to observe the precipitation stages in Fe based steels and many other alloys. Microstructural changes including precipitation are studied in age hardenable commercial alloys [23, 24]. Positron annihilation spectroscopy was used to identify defect evolution and Cr clustering in Fe-Cr alloys by various researchers [10, 15, 20, 25].

Mossbauer spectroscopy was also extensively used to study the variations in Cr distribution, viz. (i) short range ordering (ii) Cr segregation in Fe-Cr alloys by observing the variations in hyperfine field distribution [26, 27]. In this system, the preferential Cr-Cr attractive short-range interaction and Cr segregation leaves the rest of the matrix Fe rich and hence the increase in hyperfine field as compared to a random distribution of Cr in alloy.

The current study reports the positron annihilation lifetime (PAL), Transmission electron microscopy (TEM) and Mossbauer spectroscopy observations of Cr segregation in Fe-12Cr binary alloy with two initial conditions viz. (i) defect-free homogenized condition (equilibrium) (ii) quenched sub-grain microstructure (non-equilibrium). This study aims at understanding the (i) nature of precipitation (ii) role of defects in precipitation and (iii) the effect of increased Cr concentration on precipitation and the influence of high temperature $\alpha$-phase, if any, on precipitation behaviour. Corresponding results of Fe-9Cr and Fe-12Cr upon isothermal treatment are compared. These results comprising of precipitate size distribution and relative concentrations would help to compare and validate the theoretical models on solute segregation in Fe-Cr like systems.

## Experimental Procedure

Fe–12Cr (wt%) alloy was prepared using arc melting and homogenized at 1090 K for 144 h. In general, the homogenization temperature would be chosen above normalisation temperature in alloying. However, for this concentration of Cr, at corresponding temperatures, Fe-Cr alloys will be in face-centered cubic structure. Diffusion is faster and requires less activation energy in the body-centered cubic (BCC) phase than in the face-centered cubic (FCC) phase. Hence, we chose homogenization temperature of 1090 K for better mixing, despite the temperature being below normalizing temperature. The samples were cold rolled to make sheets of thickness 300 μm. The chemical composition (in wt%), as measured by Electron Probe Microanalysis (EPMA), is found to be Cr (11.66 ± 0.24) and balance Fe. No other impurities are found within the detection limits of EPMA i.e. ~ 0.01 wt%. Since carbon can form vacancy-carbon complexes [28] and metal carbides [29], carbon content is measured separately using Carbon Combustion Analyzer and is found to be around 78 ± 10 ppm which is well below the solubility limit of carbon in Fe at 300 K i.e., 450 ppm. Quenched condition is achieved by normalizing at 1423 K in a vacuum of $<10^{-6}$ Torr for 1 h followed by quenching to room temperature. These samples are subjected to isochronal heat treatment from 300 to 1223 K in steps of 50 K for 1 hr in vacuum, followed by spontaneous cooling outside the furnace. Defect-free homogenization is achieved by treating the samples at 1073 K for 2 h followed by cooling spontaneously outside furnace. This temperature choice below the initial homogenization temperature is to ensure the temperature is well below FCC loop so that FCC phase is completely ruled out during treatment. Isothermal treatments are carried out (on both quenched and defect-free homogenized samples) for different intervals at 748 K, temperature known for embrittlement in ferritic steels with Cr concentration greater than 12 wt%. Fe-9Cr samples were prepared and treated in identical conditions and the details can be found in the earlier report [20]. Cr concentration in Fe-9Cr alloy is 9.50±0.23 wt%.

PAL measurements are carried out at room temperature with a fast-fast lifetime spectrometer having a time resolution of 260 ps (FWHM) and using $^{22}$Na source. Source lifetime and intensity are measured, for subsequent corrections, using annealed Fe sample as reference. For each measurement $>10^6$ counts are accumulated and measured PAL spectra were analysed using LT program [30]. In principle, the experimental lifetime spectrum is a linear combination of n exponentials, corresponding to n distinct annihilation sites, with coefficients proportional to the fraction of positrons annihilating in that particular site. When the number of distinct annihilation sites is more than three or the lifetime variation for distinct annihilation sites is relatively less, the statistical fluctuations dominate and lifetime cannot be resolved properly. Mostly the annihilation sites at grain/sub-grain boundaries and dislocations satisfy the above conditions. We had tried fitting the experimental lifetime spectra in terms of two and three lifetime components. While the variance was lower for these fits, the lifetime values and intensities are not physically meaningful. Thus, we have resorted to only single lifetime for all the spectra measured.

TEM studies were carried out on quenched and defect-free homogenized Fe-12Cr samples after 25 hour heat treatment at 748 K. TEM studies were carried out, using a Tecnai-G2-20 TEM operating at 200 kV.The specimens for TEM study were prepared by grinding and polishing, down to 100 μm followed by electro polishing in a solution of 10% perchloric acid in methanol at ~20 V.

$^{57}$Fe Mossbauer measurements were carried out in transmission mode with $^{57}$Co(Rh) radioactive source in constant acceleration mode using standard PC-based Mossbauer spectrometer equipped with a WissEl make velocity drive. Velocity calibration of the spectrometer was done with a natural iron absorber at room temperature. As the spectra indicated a distribution of hyperfine field values, corresponding to different possible combinations of Fe, Cr atoms around $^{57}$Fe nucleus, hyperfine field distribution model was used to fit the experimental data.

## Results and Discussion

The results are discussed in four sections. In the first section, changes in positron lifetime in quenched Fe-12Cr alloy with isochronal treatment are presented, which identifies onset of precipitation stage and the subsequent dissolution at high temperature. These results are compared and discussed with reference to isochronal studies of quenched Fe-9Cr alloy [20]. Based on these results, the need for isothermal studies and the treatment temperature is outlined. In the second section, positron lifetime results of quenched Fe-9Cr and Fe-12Cr alloys, as a function of isothermal heat treatment, are presented. The differences in lifetime variations as a function of Cr concentration and, possible, associated microstructural variations are discussed. In the third section, positron lifetime results of defect-free Fe-9Cr and Fe-12Cr alloys, as a function of isothermal heat treatment, are presented and discussed the role of defects and microstructure on precipitation. In the fouth section, the observations from TEM and Mössbauer spectroscopy studies of isothermally treated Fe-12Cr alloy are presented as complementary evidence to positron results.

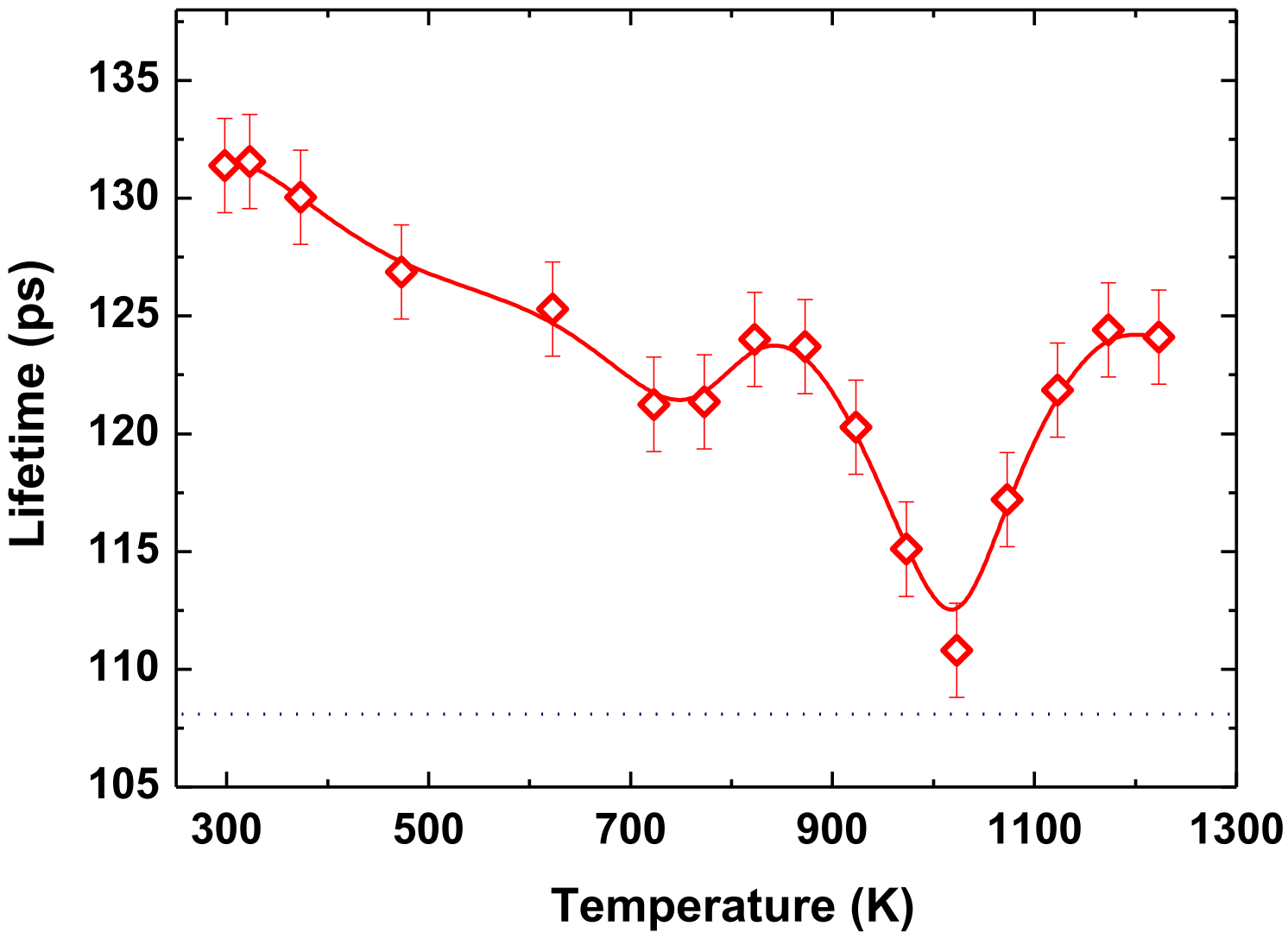

*Figure-1: Positron lifetime in quenched Fe-12Cr alloy as a function of heat treatment temperature. For reference, lifetime in defect free lattice (108 ps) is indicated with dotted line. As-quenched sample showed higher lifetime than bulk lifetime, indicating open volume defects which will be getting annealed out with increase in treatment temperature, till 723 K. Increase in lifetime above 723 K indicate fresh production of defects, in terms of precipitate-matrix interfaces, indicating precipitates formation. The precipitates dissolve back into matrix at elevated temperature, above 823 K, resulting in decreasing lifetime. The increase in lifetime above 1023 K is due to the high temperature FCC phase formation at treatment temperature, resulting in formation of defects during cooling across FCC-BCC phase transformation.*

Positron lifetime in quenched Fe-12Cr alloy as a function of isochronal heat treatment temperature is shown in figure-1. The positron lifetime variations are qualitatively identical to Fe-9Cr alloys [20]. While the lifetime of positrons in defect-free lattice of Fe-Cr is 108±2 ps, a lifetime of 131 ps is observed in quenched state. This indicates the presence of open volume defects in as-quenched state. The microstructure of quenched state is a sub-grain structure with high density of randomly distributed dislocations, similar to martensitic structure observable in martensitic steels. A microstructure with these characteristics is reported for Fe-12Cr alloy [13]. The sub-grain size is more and dislocation density is distinctly less, in this report, than expected for a quenched system. This may be due to the post-quenching heat treatment at 1003 K, which causes annealing of dislocations and growth of sub-grains due to boundary motions [13]. However, it can be considered as an indicative microstructure for the current study with regard to sub-grain structure. Indicative random dislocation structure can be understood from TEM micrograph of, identically heat treated, quenched Fe-9Cr alloy [25]. However, a small amount of the high temperature α phase, if exists, might retain as defect free grains in quenched state of Fe-12Cr alloy. This will be further discussed later. As it is understood in the case of Fe-9Cr alloy, the defects responsible for higher lifetime in quenched state of Fe-12Cr alloy are dislocations (decorated with vacancies) and sub-grain boundaries.

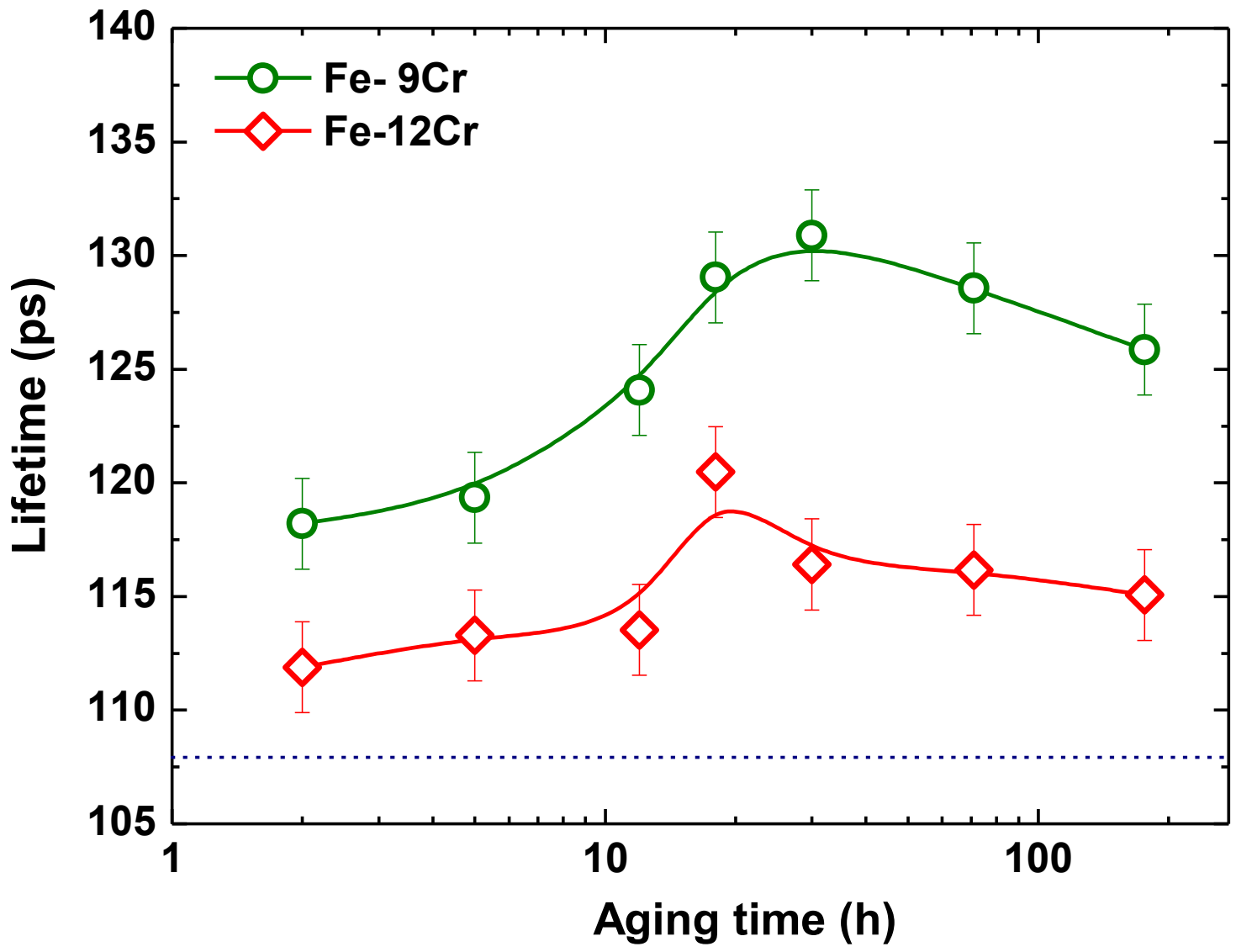

*Figure-2: Positron lifetime in quenched Fe-9Cr and Fe-12Cr alloys as a function of aging time (up to 176 h) for a treatment temperature of 748 K. Lifetime in as-quenched Fe-9Cr and Fe-12Cr alloys are 127 ps and 131 ps, respectively. The higher lifetime in Fe-9Cr over entire time range indicate relatively more open volume defect and hence more precipitate-matrix interfacial area.*

There are four stages of variation in positron lifetime between treatment temperature range of 300 K and 1273 K, namely (i) lifetime gradually decreases with temperature, as the alloy is treated at susequent higher temperatures, till 723 K, indicating the annealing of dislocations (ii) increase in lifetime above 723 K indicating formation of fresh open volume defects in terms of precipitate-matrix interfacial defects (iii) above 823 K, the lifetime decreases indicating the dissolution of precipitates into matrix (iv) the increase in lifetime above 1023 K shows that the treatment temperature is in high temperature FCC loop of phase diagram, resulting martensitic transformation across FCC-BCC phase boundary during cooling. Identical stages of positron lifetime are observed in Fe-9Cr alloy. There are, however, two noticeable changes in these stages with respect to Fe-9Cr. (a) 50 K change in onset of precipitation stage. While the onset of precipitation in Fe-12Cr alloy is 773 K, it is at 823 K in the case of Fe-9Cr alloy. This could be due to increase in Cr concentration which makes clustering kinetics relatively faster at lower temperature. (b) The lifetime during precipitation stage is larger in the case Fe-9Cr alloy (128 ps) compared to Fe-12Cr alloy (124 ps). This shows that the open volume defects at precipitate-matrix interface are relatively less in the case of Fe-12Cr alloy indicating changes in precipitate concentration and/or size. To understand these variations in precipitation characteristics isothermal studies are carried out. Isothermal heat treatment is carried out at 748 K, close to estimated onset of fresh open volume defects expected of precipitate-matrix interface, and lifetime is measured at different intervals. The temperature is chosen from lifetime variations in isochronal treatment, as it is close to the onset of precipitation. This temperature is also technologically significant due to well-known embrittlement problem at 748 K (popularly known as 475 °C embrittlement) [7].

Figure-2 shows the positron lifetime in quenched Fe-9Cr and Fe-12Cr alloys as a function of aging time at a heat treatment temperature of 748 K. Both the alloys indicate higher lifetime as compared to defect free lattice evidencing precipitation. As it is observed in isochronal treatment studies, positron lifetime in Fe-9Cr alloy is higher compared to Fe-12Cr alloy throughout the range of heat treatment. It is also observed in both the alloys that lifetime increased with aging time and then decreased beyond 30 h and 18 h in Fe-9Cr and Fe-12Cr, respectively. The increase in lifetime shows the nucleation and growth of precipitates. The crossover point could be a signature of precipitation coarsening as it will reduce the number of positron trapping sites. During coarsening the larger precipitates grow at the expense of smaller precipitates due to relatively large Cr concentration gradients at the bigger precipitates. Hence, the number density decreases and the size increase, along with increase in inter-precipitate spacing. This leads to reduction of precipitate-matrix interfacial area, and hence reduction in open volume defects, and increase in inter particle distance, combinedly reducing number of positrons trapping in open volume defects. The early crossover in Fe-12Cr, relative to Fe-9Cr also indicates that it could be due to coarsening, because, the growth is expected to be faster as the Cr concentration increases.

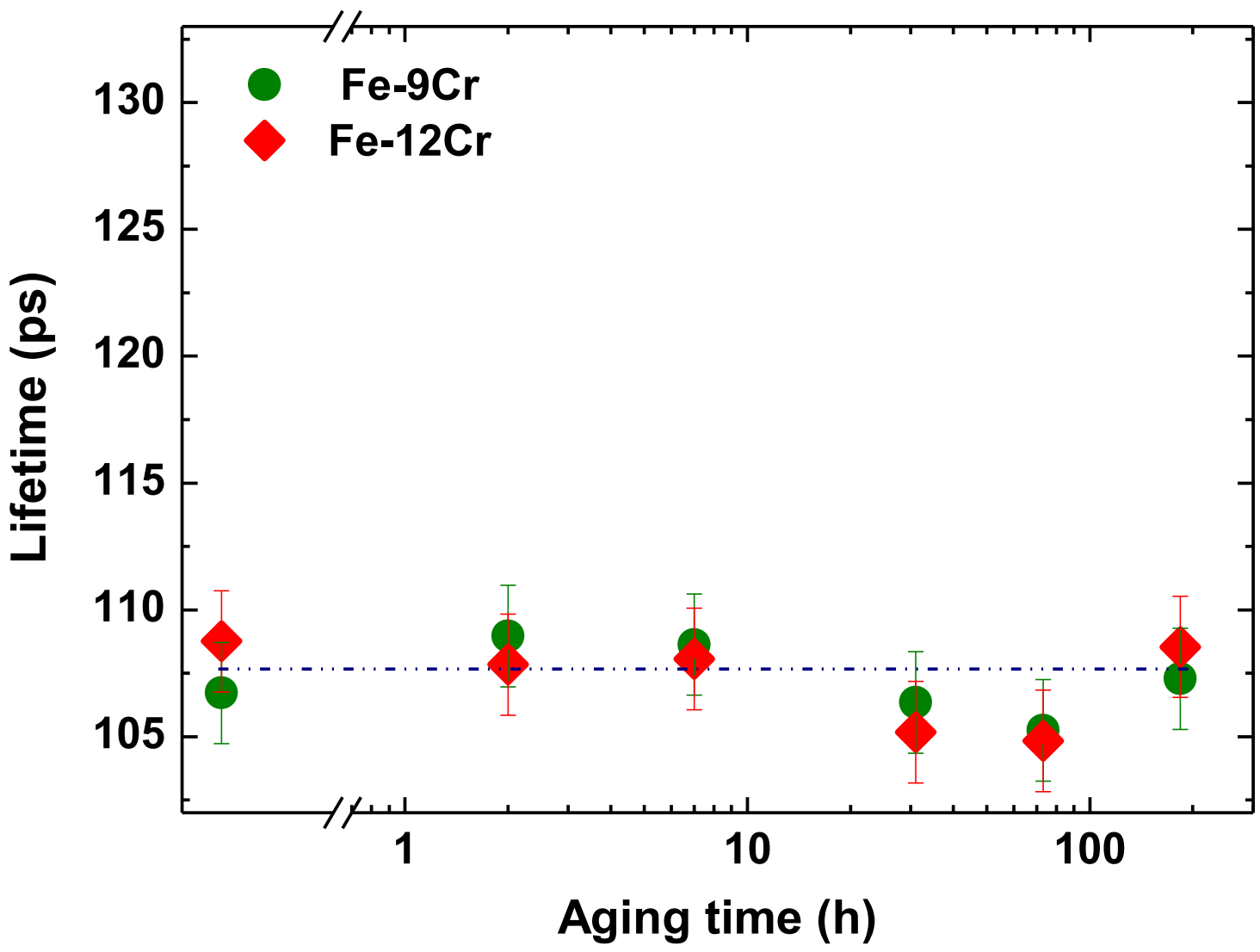

*Figure-3: Positron lifetime in defect-free homogenized Fe-9Cr and Fe-12Cr alloys as a function of aging time (up to 184 h) for a treatment temperature of 748 K. No significant variation from defect-free lattice lifetime indicates the absence of incoherent precipitates.*

The higher lifetime in Fe-9Cr alloy as compared to Fe-12Cr alloy denote two possibilities. First possibility is that the high temperature $\alpha$ phase in Fe-12Cr is significant in volume fraction. During the quenching process, this $\alpha$ phase retain at room temperature with no defects. As we have understood from Fe-9Cr alloy, defects leads to precipitation and it will also be shown in this report, when isothermal studies with defect-free initial structure is discussed. If such high temperature $\alpha$ phase exists in quenched Fe-12Cr alloy, no precipitation occurs in those defect-free grains. The second possibility is that the precipitate number density is less and the average size is more due to increase in Cr concentration in the alloy. Hence, decrease in precipitate-matrix interfacial area and larger inter-precipitate distance results in lower average lifetime in Fe-12Cr alloy as compared to Fe-9Cr alloy.

Positron lifetime in defect-free homogenized Fe-9Cr and Fe-12Cr alloys as a function of again time at a heat treatment temperature of 748 K is shown in figure-3. There is no variation in lifetime in both the alloys even after 184 hours of aging. This shows, there is no production of fresh vacancies in the sample. This indicates that incoherent precipitation is absent in defect-free alloys, highlighting the crucial role of defects on precipitation. However, there could be a segregation of Cr in the form of $\alpha'$ as aging time increases. Literature suggests such attractive potentials between Cr-Cr for alloy with Cr concentration above ~9%, though their kinetics is very slow. However, we will not be able to identify that using positron lifetime spectroscopy as it will not create any defects, being coherent with matrix. Also, it would be difficult to identify even in TEM because of the less elemental contrast between Fe and Cr for attenuation of electrons. However, Mössbauer spectroscopy is capable of identifying such segregation, but did not find any evidence, at least till 25 hours of heat treatment. This is discussed in the later part of this manuscript.

Dependence of initial heat treatment, whether it is at a temperature within the FCC loop or at a temperature below the loop, on microstructure and subsequent precipitation on heat-treatment is shown in both Fe-9Cr and Fe-12Cr alloys. This study indicates the variation in microstructure, as observed through open volume defects and TEM observations, due to quenching across phase boundary. In the process, this is the only study which explicitly describe the importance of starting microstructure in carbon-free Fe-Cr binary alloys. In the process, this work also reported bulk lifetime, pertaining to defect-free lattice, of Fe-(9, 12)Cr alloys, for the first time.

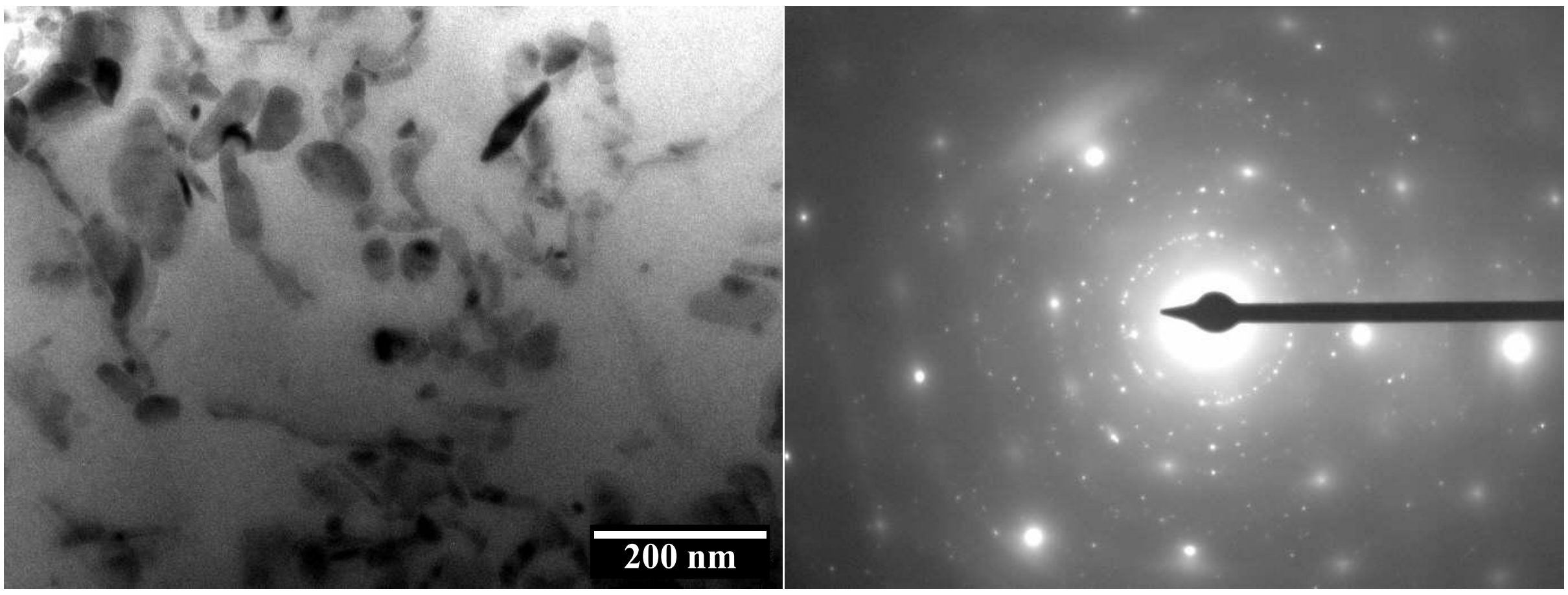

*Figure-4: TEM micrograph of quenched Fe-12Cr alloy heat treated at 748 K for 25 h showing distribution of precipitates. The precipitates are quite larger in size as compared to the precipitates observed in Fe-9Cr alloy in the previous study [20]. However, the heat treatments are different in both the cases.*

To understand the precipitation behaviour due to isothermal treatment, TEM studies are carried out on quenched and defect-free homogenized Fe-12Cr alloys after 25 hours of treatment. Figure-4(a) shows the TEM micrograph of quenched Fe-12Cr alloys subsequently heat treated at 748 K for 25 hours. Precipitates, mostly elongated in shape, are present throughout the sample. A few regions were seen with retained dislocations. Whenever such dislocation structure is seen, precipitates are absent in those sub-grains. The size distribution of these precipitates is quite larger, 30-100 nm, against what it was observed in Fe-9Cr alloy, < 10 nm. However, there is a difference in treatment time interval between these two measurements for which the TEM studies were carried out. Figure-4(b) shows the selected area diffraction pattern of a cluster of precipitates. The d-spacing values, distinct from BCC Fe-Cr phase, measured from diffraction are 2.46±0.10, 2.16±0.10 and 1.51±0.10 Å. According to JCPDS data, all these value are consistent with σ-phase of Fe–Cr, indicating the precipitates are indeed Fe-Cr intermetallic phase, with tetragonal crystal structure. These observations indicate a possible explanation for contradiction between theory and experiments regarding the nature of phase responsible for embrittlement. The lattice parameter differences are very small (<0.04 Å) between $\alpha$ phase and chromium rich $\alpha'$ phase, the line defects movement is hardly hindered by these precipitates [8]. Despite this knowledge, for a very long time, the $\alpha'$ phase is ascribed as a reason for hardening. This work brings new evidence towards the possible presence of σ-phase in Fe-(>9%)Cr based ferritic/martensitic steels. This work stands alone with regard to σ-phase identification in Fe-Cr alloys with Cr percentage as low as 9%.

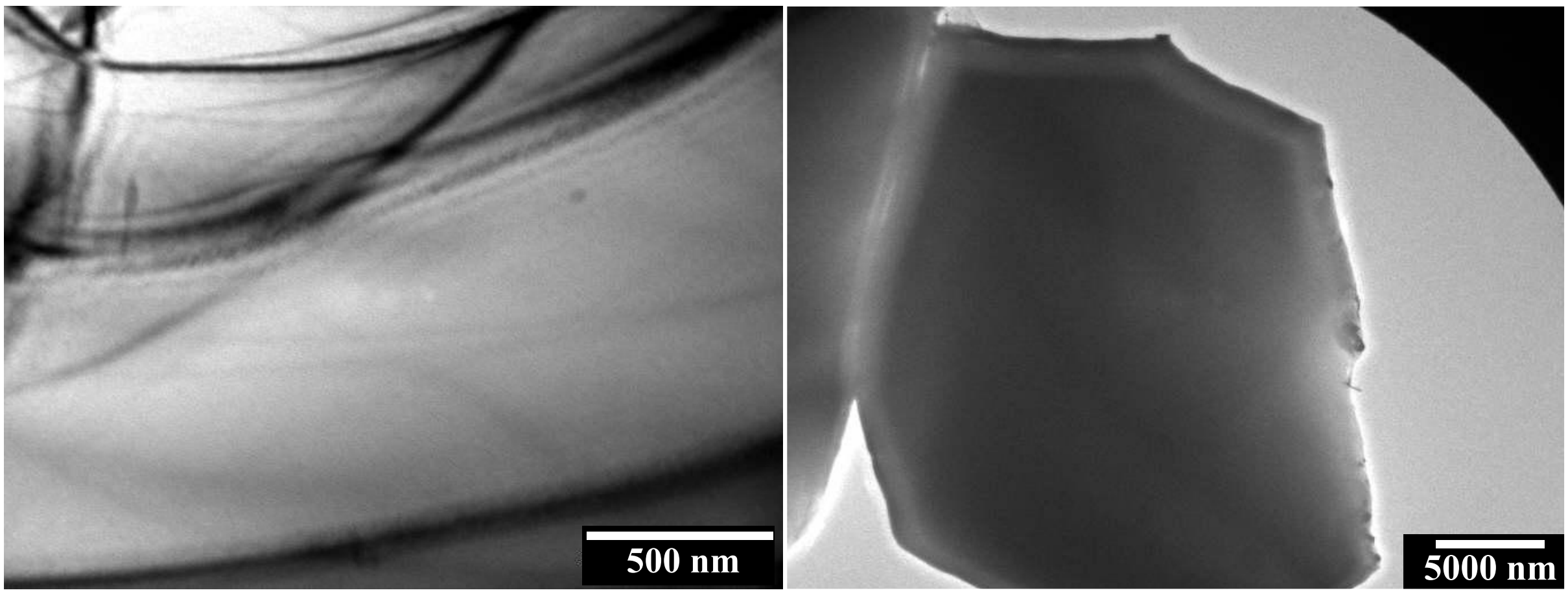

*Figure-5: TEM micrograph of defect-free homogenized Fe-12Cr alloy heat treated at 748 K for 25 h. (a) No precipitates are observed complementing the results of positron lifetime which shows no changes in lifetime with heat treatment. (b) Image of a single grain captured in low resolution mode to depict the typical grain size.*

With regard to coexistence of $\alpha$-phase during normalizing treatment, if there is a significant high temperature $\alpha$-phase, during quenching there is no phase transition for this phase and these specific grains might result in defect-free as well as precipitate-free phase in the isothermally treated alloy. However, no such phase is identified in TEM studies. It is either very negligible in volume fraction or the grains must be small enough to be altered by surrounding high temperature $\gamma$-phase volume expansion during quenching.

Figure-5(a) show the TEM micrograph of defect-free homogenized Fe-12Cr alloy subsequently heat treated at 748 K for 25 hours. The observations show that the matrix is precipitate-free and dislocation free, throughout the sample. There is no sub-grain structure observed in this sample, while the grain size is of the order of 20 μm (figure-5(b)). These observations are consistent with the understanding that the Cr segregation is induced by defects. This is consistent with regard to precipitate embrittlement in ferriritc/martensitic alloys as they contain large dislocation density resulted in their making process.

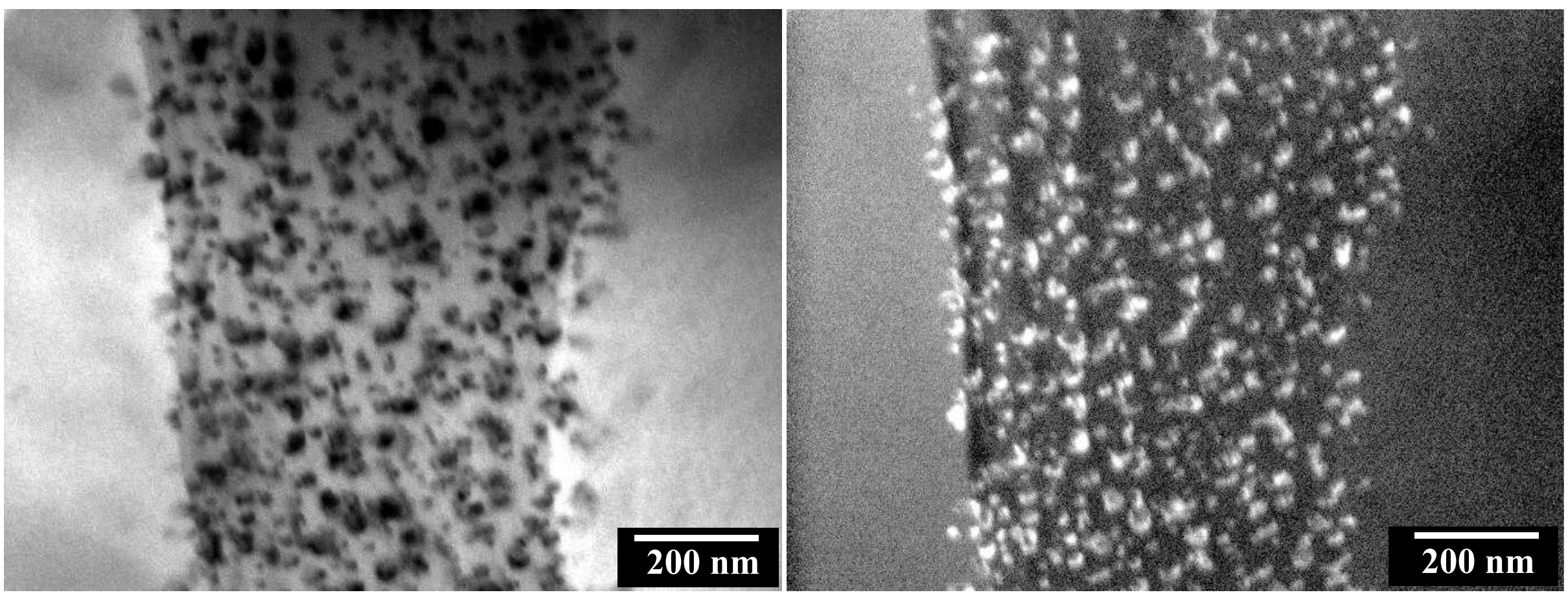

*Figure-6: TEM micrographs of defect-free homogenized Fe-12Cr alloy heat treated at 748 K for 25 h. (a) bright field and (b) dark field micrographs showing a grain boundary decorated with oriented precipitates.*

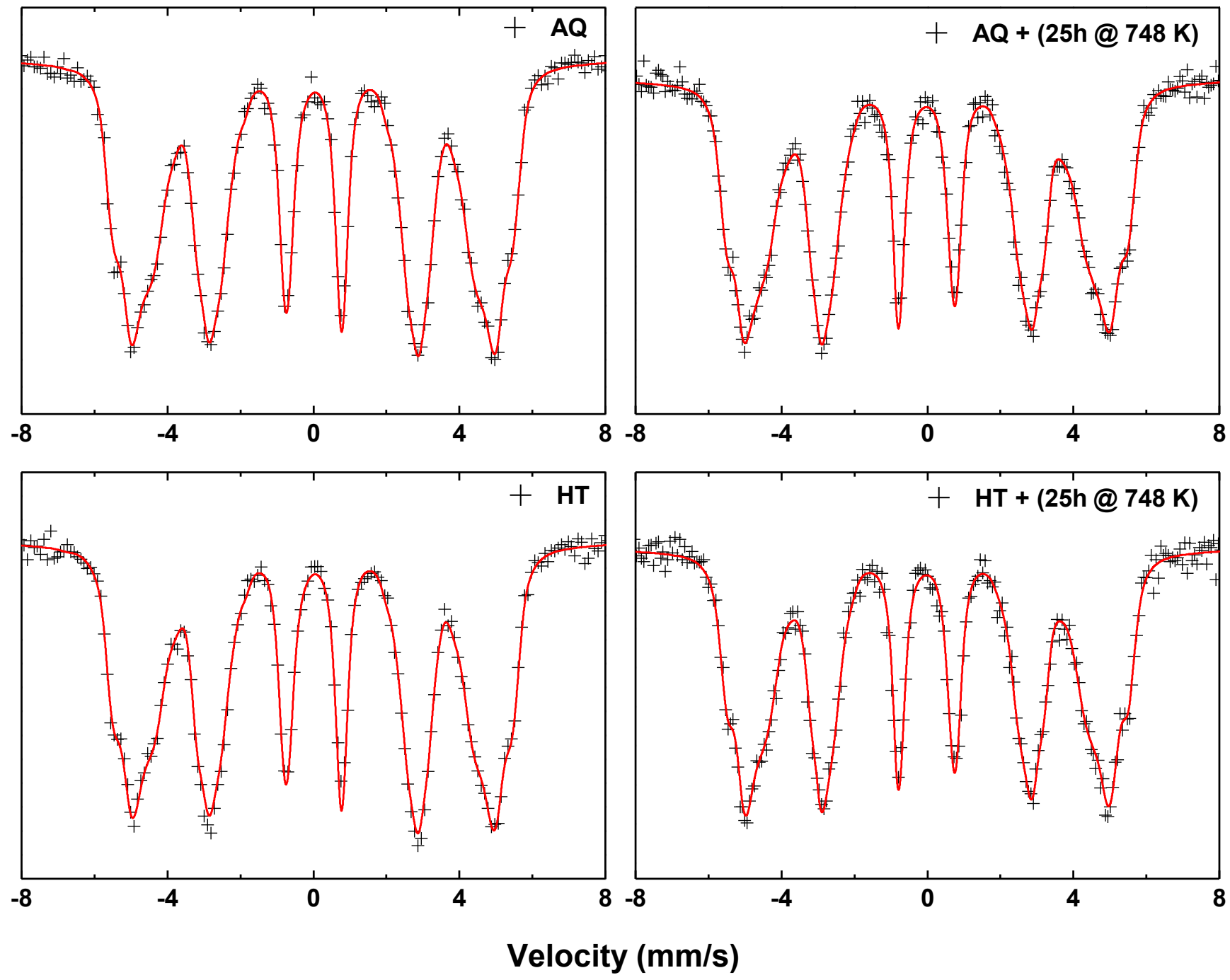

*Figure-7: Mossbauer spectra of Fe-12Cr alloy recorded in quenched and defect-free homogenized conditions and the corresponding alloys after 25 h heat treatment at 748 K. The spectrum shows broad peaks. A spectrum of this kind indicates a distribution of hyperfine field due to different possible combinations of atoms and their position around $^{57}$Fe isotope.*

However, highly oriented precipitates are seen decorating grain boundaries. Figure-6 shows the bright field and dark field micrographs showing oriented precipitates along a grain boundary. Since these grain boundaries are negligible in fraction compared to the grain volume and positron diffusion length in metals and alloys (~100 nm) is too small compared to the size of grains, the contribution of grain boundary defects to positron lifetime are considered negligible. Nevertheless, this observation that grain boundaries alone contain precipitates in dislocation free sample shows that the open volume defects mediate Cr diffusion leading to clustering and σ-phase formation. It should be noted that the sub-grain boundaries in Fe-9Cr and Fe-12Cr alloys too are decorated with precipitates more densely than the density within the sub-grains.

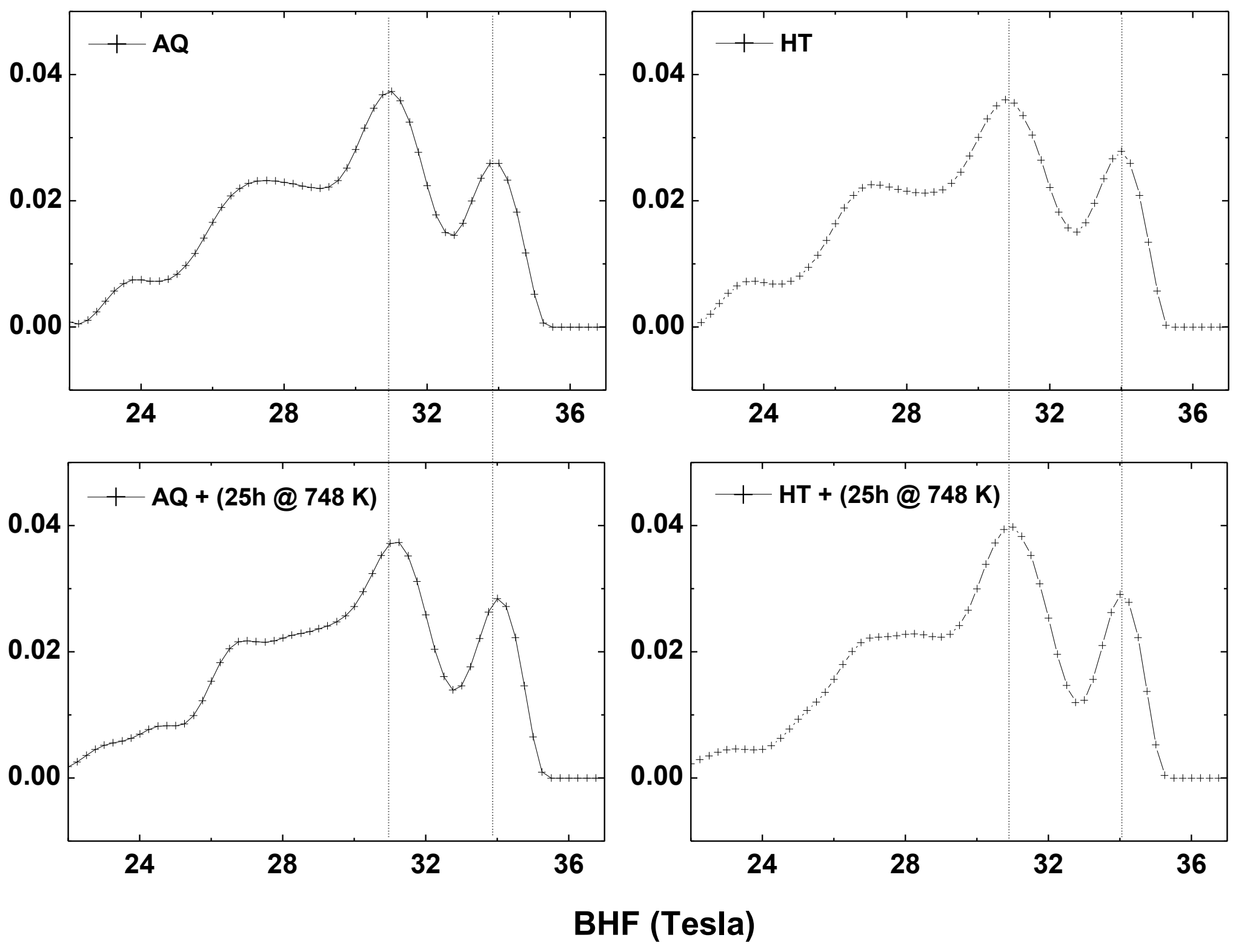

*Figure-8: Distribution of magnetic hyperfine field in Fe-12Cr alloy in quenched and defect-free homogenized conditions and the corresponding alloys after 25 h heat treatment at 748 K. After heat treatment, shift in hyperfine field towards higher values is observed in the case of quenched sample. No significant change observed in the case of defect-free homogenized sample.*

Figure-7 shows the $^{57}$Fe Mössbauer spectra of Fe-12Cr alloy in both as-prepared forms (i) quenched (ii) defect-free homogenized and the respective sample after 25 hours of heat treatment. Mössbauer spectra of investigated alloys take a shape typical of Zeeman sextet pattern. The lines are broad indicating that it originates from atoms being subjected to distributed magnetic hyperfine fields. The spectra are similar to the one reported, for similar composition, with regard to gross features of peaks [31]. The spectra are fitted using hyperfine field distribution subspectrum method. Figure-8 shows the fitted values of hyperfine field distribution in corresponding samples. The results show that the field distribution in quenched alloy shifted towards higher $B_{hf}$ values marginally, indicating that the matrix is enriched with Fe atoms indicating the Cr clustering. However, in defect-free homogenized sample, the distribution of hyperfine fields remains same before and after heat treatment. While it was argued in the earlier discussions that positron annihilation lifetime and TEM are incapable identifying $\alpha'$-phase inside $\alpha$-phase, being coherent and their respective electron beam attenuation being similar, Mössbauer could able to identify the $\alpha'$-phase formation with change in hyperfine field of rest of the matrix. However, the observation of no change in hyperfine fields show absence of significant $\alpha'$ phase in defect-free homogenized sample, confirming the role of defects on accelerated precipitation, in quenched samples, at this concentration level. A varying length of aging time may be of interest for future studies to see what happens for prolonged periods. The observations from Mössbauer studies complemented the positron annihilation lifetime and TEM studies by confirming the changes in overall matrix chemical composition which is not possible to show from the latter two studies. This evidence shows the Cr rich nature of precipitates.

## Conclusions

Positron lifetime, TEM and Mössbauer spectroscopy studies were carried on Fe-12Cr alloy to understand the precipitation behaviour as a function of heat treatment and the following observations were brought out.

From isochronal studies of quenched alloy, it is found that precipitates nucleate at a temperature 50 K lower than in the case of Fe-9Cr alloy. Isothermal studies showed that for a given temperature the precipitates grow faster in the case of Fe-12Cr alloy and their size is larger while number density is less, compared to Fe-9Cr alloy. Diffraction pattern confirmed them as tetragonal intermetallic precipitates. Mössbauer studies showed enhancement of hyperfine field values proving that the precipitates are enriched with Cr.

Isothermal studies on defect free alloy showed no change in positron lifetime indicating no precipitation. Along with positron annihilation, TEM studies also confirmed the absence of intermetallic precipitation. Further, Mössbauer spectroscopy studies confirmed Cr remain soluble (at least up to 25 hours of aging), indicating even BCC $\alpha'$ segregation is a very slow process in the absence of defects, again confirming the crucial role of dislocations/defects on precipitation. Furthermore, the precipitation along grain boundaries strengthened the idea of defect mediated precipitation.

Experimental lifetime of positron in defect-free lattice of Fe-9Cr and Fe-12Cr is found to be 108±2 ps. This serve as a reference to identify the sample initial state as it is significantly changing the microstructure in further treatments. The capability of distinguishing the initial state will help better comparison of corresponding experimental and theoretical results. Literature suggests that it was never tried to incorporate the dislocations and sub-grain boundaries in theoretical models to check the influence of them on Cr precipitation.

## Acknowledgements

Authors thank Dr. R. Rajaraman and Dr. G. Amarendra for useful discussions, Dr. R. Govindraj for help in making samples, Ms. Shabana Khan for help in TEM sample preparation, Dr. V. Chandramouli and Dr. C. Sudha for carbon analysis and EPMA analysis, respectively. Thanks to Dr. Ajay Gupta and Dr. V. Ganesan for their constant support and encouragement. One of the authors (SHB) acknowledges DAE, India for research fellowship.

## References

[1] E. Lucon, P. Benoit, P. Jacquet et.al., *Fus. Eng. Des.* **81**, 917-923 (2006).
[2] S. L Manan, S.C. Chetal, B. Raj, S.B. Bhoje, *Trans. Ind. Inst. Metals* **56**, 155-178 (2003).
[3] D. S. Gelles unpublished work through R. L. Klueh, ORNL/TM-2004/176.
[4] H. M. Chung, B. L. Loomis and D. L. Smith, "U.S. Contribution, 1994 Summary Report, Task T12: Compatibility and Irradiation Testing of Vanadium Alloys, " ed. D. L. Smith, ANL/FPP/TM-287, ITER/US/95/IV MAT 10, March 1995, pp 147-155.
[5] Yusuke Kudo, Kouichi Kikuchi, Masakatsu Saito, J. Nucl. Mater. **307-311**, 471 (2002).
[6] M. Rieth, M. Schirra, A. Falkenstein, P. Graf, S. Heger, H. Kempe, R. Lindau, H. Zimmermann, Karlsruhe Research - Scientific reports - **FZKA 6911** (2003).
[7] T. Angeliu, E. L. Hall, M. Larsen, A. Linsebigler, C. Mukira, Metall. Mater. Trans. A **34A**, 927 (2003).
[8] L. Malerba , A. Caro, J. Wallenius, J. Nucl. Mater. **382**, 112 (2008).
[9] Wei Xiong, Malin Selleby, Qing Chen, Joakim Odqvist, Yong Du, Crit. Rev. Solid. State **35**, 125 (2010).
[10] M. Lambrecht and L. Malerba, Acta Mater. **59**, 6547 (2011).
[11] I. Mirebeau and G. Parette, Phys. Rev. B **82**, 104203 (2010).
[12] C. Heintze, A. Ulbricht, F. Bergner and H. Eckerlebe, J. Phys.: Conf. Ser. **247**, 012035 (2010).
[13] M. Matijasevic and A. Almazouzi, J. Nucl. Mater. **377**, 147 (2008).
[14] T. Ezawa, T. Akashi and R. Oshima, J. Nucl. Mater. **212–215**, 252 (1994).
[15] J. Kwon, T. Toyama, Y.-M. Kim, W. Kim and J.-H. Hong, J. Nucl. Mater. **386–388**, 165 (2009).
[16] P. Olsson, C. Domain and J. Wallenius, Phys. Rev B **75**, 014110 (2007).
[17] G. Bonny, D. Terentyev, L. Malerba and D. Van Neck, Phys. Rev. B **79**, 104207 (2009).
[18] D. Terentev, P. Olsson, L. Malerba and A.V. Barashev, J. Nucl. Mater. **362**, 167 (2007).
[19] S. Lu, Q.-M. Hu, R. Yang, B. Johansson and L. Vitos, Phys. Rev. B **82**, 195103 (2010).
[20] S. Hari Babu, R. Rajaraman, G. Amarendra, R. Govindaraj, N. P. Lalla, Arup Dasgupta, Gopal Bhalerao, C. S. Sundar, Philos. Mag. **92**, 2848 (2012).
[21] P. Hautojärvi, Positrons in solids, Springer, Berlin, 1979.
[22] Krause Rehburg, H. S. Leipner, Positron annihilation in semiconductors: Defect studies, Springer-Verlag, Berlin, 1998.
[23] R Ferragut, A Somoza, A Dupasquier, J. Phys.: Condens. Matter. **8**, 8945 (1996).
[24] A. Dupasquier, R. Ferragut, M. M. Iglesias, M. Massazza, G. Riontino, P. Mengucci, G. Barucca, C. E. Macchi, A. Somoza, Phil. Mag. A **87**, 3297 (2007).
[25] S. Hari Babu, Positron Annihilation Studies of Model Fe-Cr Alloys and Ferritic/Martensitic Steels, PhD Thesis, HBNI, Mumbai, 2013.
[26] S. M. Dubiel, J. Cieslak, Phy. Rev. B **83**, 180202(R) (2011).
[27] H. Kuwano, Trans. Jpn. Inst. Met. **26**, 473 (1985).
[28] P. Hautojarvi, J. Johansson, A. Vehanen, J. Yli-Kauppila and P. Moser, Phys. Rev. Lett. **44**, 1326 (1980).
[29] S. Hari Babu, K. V. Rajkumar, S. Hussain, G. Amarendra, C. S. Sundar and T. Jayakumar, J. Nucl. Mater. **432**, 266 (2012).
[30] J. Kansy, *Nucl. Instr. and Methods A* **374**, 235-244 (1996).
[31] S. M. Dubiel, J. Cieslak, Mater. Lett. **107**, 86 (2013).